\documentclass[seceq]{ptptex}

\newcommand{\ket}[1]{|{#1}\rangle}
\newcommand{\bra}[1]{\langle {#1}|}
\newcommand{\ubexav}[2]{{\underline a}\sp{#1}\sb{#2}}
\newcommand{\ubexv}[2]{{\underline v}\sp{#1}\sb{#2}}
\newcommand{\axialjc}[2]{j\sp{#1}\sb{A{#2}}}
\newcommand{\vectorjc}[2]{j\sp{#1}\sb{V{#2}}}
\newcommand{\diff}[2]{\frac{\delta {#1}}{\delta {#2}}}
\newcommand{\fpi}{f\sb{\pi}}
\newcommand{\mpi}{m\sb{\pi}}
\newcommand{\chrf}{$\chi$RF}
\newcommand{\smatrix}{\hat{{\cal S}}}
\newcommand{\smatrixon}{\hat{\boldsymbol{S}}}

\title{Off-Shell Extension of the Chiral Reduction Formula}

\author{Hiroyuki~\textsc{Kamano}}

\inst{Department of Physics, Osaka University, Toyonaka, Osaka 560-0043, Japan}

\abst{
We present a proper off-mass-shell extension of 
the chiral reduction formula ({\chrf}) proposed by Yamagishi and Zahed. 
This is achieved by rewriting the {\chrf} in a
manifestly consistent form with the conventional LSZ reduction formula.
}

\begin{document}

\maketitle
\section{Introduction}
\label{sec1}
There is no doubt that chiral symmetry is important for a theoretical
understanding of the low energy properties of QCD.
The spontaneous, and also explicit breaking of this symmetry 
governs various aspects of hadronic processes and plays a crucial role
in the construction of effective models.

On the basis of chiral symmetry, Yamagishi and Zahed developed
a general framework for analyzing pion-induced hadronic processes\cite{Yam96}.
This framework consists of two master equations 
for the extended S-matrix, which are based on the gauge covariant
divergence equations proposed by Veltman and Bell\cite{Vel66,Bell67}.
Each master equation represents transformation properties of 
the extended S-matrix under the local vector and axial transformations,
subject to the following asymptotic conditions
for the axial current: 
\begin{equation}
A\sb{\mu}\sp{a}(x) 
\rightarrow 
-\fpi\partial\sb{\mu}\pi\sp{a}\sb{{\rm in,out}}(x)
+ \cdots
,\ \
(x\sp{0} \rightarrow \mp\infty)
\label{eq11}
\end{equation}
and
\begin{equation}
\partial\sp{\mu}A\sb{\mu}\sp{a}(x) 
\rightarrow \fpi\mpi\sp{2}\pi\sp{a}\sb{{\rm in,out}}(x)
+ \cdots
.\ \
(x\sp{0} \rightarrow \mp\infty)
\label{eq12}
\end{equation}
One noteworthy result obtained within this framework is that
the master equation for the axial transformation provides
a new reduction formula, the chiral reduction formula ({\chrf}),
for scattering amplitudes involving any number of pions with their 
\textit{physical} masses.
With this formula, the relevant on-shell Ward identities 
satisfied by those amplitudes can be derived. 
These identities are expressed in terms of the Green functions of 
well-defined current and density operators,
and exactly embody the consequences of the broken $SU(2)$$\times$$SU(2)$ 
chiral symmetry without the need for any model or expansion scheme.
A number of investigations based on the {\chrf}
have been carried out for hadron reactions 
in the resonance region~\cite{Ste97,Lee99,Kam05-1}
and hadronic matter~\cite{Ste96,Ste97-2,Lee98,Dus06}.
These investigations have demonstrated that the {\chrf} 
is a powerful tool for clarifying the role of chiral symmetry 
in those processes in a model-independent way.

To this time, there have been no studies aimed of determining
how amplitudes behave in the off-shell region
within the framework of Ref.~\citen{Yam96}. 
In principle, the {\chrf} can be applied not only to scattering amplitudes
but also to internal vertex functions
and potentials in which the attached pion legs can be off the mass shell.
Although the off-shell amplitudes themselves are artificial, and
not directly related to physical quantities, 
such amplitudes are required
in some theoretical calculations.
For instance, in scattering problems,
the on-shell scattering amplitudes
are calculated by using off-shell potentials 
(in the Lippmann-Schwinger equation, etc.).
This situation, i.e. that the off-shell amplitudes are employed 
in calculations of observables, is also found in many-body problems.
Thus in attempting to treat these problems on the basis of the {\chrf},
it is necessary to make clear the off-shell structure.

However, care must be taken in applying the {\chrf} 
presented in Ref.~\citen{Yam96} 
to off-shell pions, because this formula was originally constructed 
for the purpose of treating the on-shell pions.
Indeed, in contrast to on-shell cases, we find that 
its naive application is invalid off the mass shell.
The purpose of this paper is to carefully examine 
the derivation of {\chrf}
and to formulate a proper method of extending  
the {\chrf} off the mass shell. 

This paper is organized as follows.
In \textsection\ref{sec2}, we briefly review the theoretical framework
developed in Ref.~\citen{Yam96}.
In \textsection\ref{sec3}, we explain in detail the kind of 
interpolating pion field chosen 
within the framework of Ref.~\citen{Yam96}. 
Generally, the interpolating pion field must be defined 
in order to relate scattering amplitudes to
the Green functions of current and density operators.
This is also the case for the {\chrf}.
The problem inherent in the {\chrf} will become clear through 
this explanation, because the off-shell behavior of amplitudes 
is fixed once an interpolating pion field is given.
In \textsection\ref{sec4}, we present an off-shell extension of 
the {\chrf} that is consistent with the framework of Ref.~\citen{Yam96}.
A summary is given in \textsection\ref{sec5}.
\section{Chiral reduction formula}
\label{sec2}
A fundamental quantity in the theoretical framework developed in
Ref.~\citen{Yam96} is the extended S-matrix,
$\smatrix = \smatrix[v\sb{\mu}\sp{a},a\sb{\mu}\sp{a},s,J\sp{a}]$,
which is a functional of the vector, axial vector, scalar 
and pseudoscalar external fields.
Then, the current and density operators 
$\hat{\cal O} = 
( \vectorjc{a}{\mu}, \axialjc{a}{\mu}, \fpi\hat{\sigma}, \hat{\pi}\sp{a} )$
conjugate to the corresponding external fields
$\phi = ( v\sp{a}\sb{\mu}, a\sp{a}\sb{\mu}, s, J\sp{a} )$
are defined as
\begin{equation}
\hat{\cal O}(x) = -i\smatrix\sp{\dag}\diff{\smatrix}{\phi(x)}.
\label{eq21}
\end{equation}
Here, $\hat{\pi}\sp{a}(x)$ is the pseudoscalar-isovector density operator,
and $\hat{\sigma}(x)$ is the scalar-isoscalar density operator from which 
the pion decay constant $\fpi$ has been subtracted.
In the presence of the external fields,
the operators $\vectorjc{a}{\mu}$, $\axialjc{a}{\mu}$ and $\hat{\pi}\sp{a}$
are related to the ordinary vector and axial currents as
\begin{equation}
V\sp{a}\sb{\mu}(x) =  \vectorjc{a}{\mu}(x) 
                    + \fpi\ubexav{ab}{\mu}(x) \hat{\pi}\sp{b}(x)
\label{eq22}
\end{equation}
and
\begin{equation}
A\sp{a}\sb{\mu}(x) =  \axialjc{a}{\mu}(x)
                    + \fpi\sp{2}a\sb{\mu}\sp{a}(x)
                    - \fpi\nabla\sp{ab}\sb{\mu}\hat{\pi}\sp{b}(x),
\label{eq23}
\end{equation}
respectively, where we use the notation
$\ubexav{ac}{\mu} \equiv \varepsilon\sp{abc}a\sp{b}\sb{\mu}$ and
$\nabla\sp{ac}\sb{\mu} \equiv \delta\sp{ac}\partial\sb{\mu} + \ubexv{ac}{\mu}$.
Owing to the Bogoliubov causality condition~\cite{Bog}, 
the $T\sp{\ast}$ product of the operators $\hat{\cal O}$ 
can be expressed as
\begin{equation}
T\sp{\ast}(\hat{\cal O}(x\sb{1})\cdots\hat{\cal O}(x\sb{n}))
= 
(-i)\sp{n}\smatrix\sp{\dag}
\diff{}{\phi(x\sb{1})} \cdots \diff{}{\phi(x\sb{n})} \smatrix.
\label{eq24}
\end{equation}

The essence of the framework presented in Ref.~\citen{Yam96} is embodied by
two linear master equations for the extended S-matrix,
\begin{equation}
T\sb{V}\sp{a}(x) \smatrix = 0,
\label{eq25}
\end{equation}
\begin{equation}
T\sb{A}\sp{a}(x) \smatrix = 0,
\label{eq26}
\end{equation}
where
\begin{equation}
T\sb{V}\sp{a}(x) =
\left(
  \nabla\sb{\mu}    \diff{}{v\sb{\mu}}
+ \ubexav{}{\mu}    \diff{}{a\sb{\mu}}
+ {\underline J}    \diff{}{J}
\right)\sp{a}(x),
\label{eq27}
\end{equation}
\begin{eqnarray}
T\sb{A}\sp{a}(x)
&=&
\left[
- (\Box + \mpi\sp{2} + K)\diff{}{J} 
+ iJ
+ \frac{1}{\fpi} X\sb{A}
- \frac{1}{\fpi} 
  \left( \nabla\sp{\mu}a\sb{\mu} - \frac{J}{\fpi} \right) \diff{}{s}
\right]\sp{a}(x)
\label{eq28}
\end{eqnarray}
and
\begin{equation}
K\sp{ab}(x) =
( \nabla\sp{\mu}\nabla\sb{\mu} - \ubexav{\mu}{}\ubexav{}{\mu}
+ s - \Box 
)\sp{ab}(x),
\ \
X\sb{A}\sp{a}(x) = 
\left(
  \nabla\sb{\mu}   \diff{}{a\sb{\mu}}
+ \ubexav{}{\mu}   \diff{}{v\sb{\mu}}
\right)\sp{a}(x).
\label{eq29}
\end{equation}
In Eqs.~(\ref{eq27})-(\ref{eq29}), the isospin indices with respect to which 
the contraction is taken are suppressed.
The operator $T\sb{V}\sp{a}(x)$ is the (local) isospin generator, 
and hence Eq.~(\ref{eq25}) represents the isospin invariance 
of the extended S-matrix.
Equation~(\ref{eq26}) represents the transformation properties of 
$\smatrix$ under axial transformations. 
By applying the functional derivatives of $\phi(x)$ 
to the master equations and using Eqs.~(\ref{eq21}) and~(\ref{eq24}), 
we straightforwardly obtain the vector and axial Ward identities
satisfied by the Green functions of $\hat{\cal O}(x)$.

A crucial point in the derivation described above is that
the asymptotic conditions (\ref{eq11}) and (\ref{eq12})
are exactly incorporated in the master equations
without the need for any specific model or expansion scheme.
Consequently, the resulting Ward identities incorporate
not only the spontaneous chiral symmetry breaking,
but also its explicit breaking to all orders in the quark masses. 
This property of the master equations allows us 
to derive the Ward identities 
for scattering amplitudes involving pions with their 
\textit{physical} masses.
This contrasts with the situation in
other theoretical frameworks, 
such as the chiral perturbation theory, 
in which the pion mass is expressed as a
perturbative extrapolation from the chiral symmetric point.
The Ward identities are integrated into 
the form of a reduction formula,
called the chiral reduction formula ({\chrf}).

We now explain the {\chrf} in detail.
First, we note that
the pseudoscalar-isovector density 
$\hat{\pi}\sp{a}(x)$ possesses the asymptotic form 
$\hat{\pi}\sp{a}(x) \rightarrow \pi\sp{a}\sb{{\rm in,out}}(x)\ 
(x\sp{0} \rightarrow \mp\infty)$, 
which follows from the construction of the master equations
and the asymptotic conditions (\ref{eq11}) and (\ref{eq12}).
This fact enables us to identify $\hat{\pi}\sp{a}(x)$ as the (normalized) 
interpolating pion field.
Then, through this identification, 
the axial Ward identities satisfied by the Green functions of 
$\hat{\cal O}(x)$ can be related to
scattering amplitudes including pions.
The identification is emphasized by formally solving 
the axial master equation (\ref{eq26}) 
for $\delta \smatrix / \delta J\sp{a}$
under the condition
$\hat{\pi}\sp{a}(x) \rightarrow \pi\sp{a}\sb{{\rm in,out}}(x)\ 
(x\sp{0} \rightarrow \mp\infty)$. 
The result is\footnote{In Ref.~\citen{Yam96}, 
Eq.~(\ref{eq210}) is called the (axial) master equation.}
\begin{eqnarray}
\diff{\smatrix}{J\sp{a}(x)} &=& 
 i \smatrix \pi\sb{\rm in}\sp{a}(x)
+i \smatrix \int d\sp{4}y
   G\sb{R}\sp{ab}(x,y)\left(K\pi\sb{\rm in}\right)\sp{b}(y)
\nonumber\\
&&
+ \int d\sp{4}y G\sb{R}\sp{ab}(x,y)
  \left[
       -i J
       + \frac{1}{\fpi}
         \left( 
         \nabla\sp{\mu}a\sb{\mu} - \frac{J}{\fpi}
         \right) \diff{}{s}
       - \frac{1}{\fpi} X\sb{A}
  \right]\sp{b}(y)  \smatrix
\nonumber\\
&=& 
 i \pi\sb{\rm in}\sp{a}(x) \smatrix
+i \int d\sp{4}y 
   G\sb{A}\sp{ab}(x,y)\left(K\pi\sb{\rm in}\right)\sp{b}(y) \smatrix
\nonumber\\
&&
+ \int d\sp{4}y G\sb{A}\sp{ab}(x,y)
  \left[
       -i J
       + \frac{1}{\fpi}
         \left( 
         \nabla\sp{\mu}a\sb{\mu} - \frac{J}{\fpi}
         \right) \diff{}{s}
       - \frac{1}{\fpi} X\sb{A}
  \right]\sp{b}(y)  \smatrix .
\nonumber\\
&&
\label{eq210}  
\end{eqnarray}
Here, $G\sb{R}$ and $G\sb{A}$ are the retarded and advanced Green functions, 
respectively, which satisfy
\begin{equation}
(-\Box - \mpi\sp{2} - K)\sp{ab}(x)
G\sp{bc}\sb{R,A}(x,y)
=\delta\sp{ac}\delta\sp{(4)}(x-y).
\label{eq211}
\end{equation}
Using Eq.~(\ref{eq210}), we can directly derive 
the commutation relations for the creation and annihilation operators
of the pion with the extended S-matrix $\smatrix$.
We find
\begin{equation}
[ a\sp{a}\sb{{\rm in}}(k), \smatrix ] = R\sp{a}(k)\smatrix, 
\ \ \ \  
[ \smatrix , a\sp{a\dag}\sb{{\rm in}}(k) ] = R\sp{a}(-k)\smatrix,
\label{eq212}
\end{equation}
where 
\begin{eqnarray}
R\sp{a}(k) &=&
\int d\sp{4}x e\sp{+ikx}
\left[
 iJ
+ \frac{1}{\fpi} X\sb{A}
- K\diff{}{J} 
- \frac{1}{\fpi} 
  \left( \nabla\sp{\mu}a\sb{\mu} - \frac{J}{\fpi} \right) \diff{}{s}
\right]\sp{a}(x)
\label{eq213}.
\end{eqnarray}
Note that we have rewritten the r.h.s. of the commutation relations 
(\ref{eq212}) in forms that do not involve the asymptotic pion field.
[Compare these with Eqs.~(6.2) and (6.3) in Ref.~\citen{Yam96}.]

Iterative use of Eq.~(\ref{eq212}) yields
the {\chrf} for scattering amplitudes 
involving any number of pions with their physical masses,\footnote{ 
We consider the case in which no two pions have equal momenta.}
\begin{eqnarray}
&&
\bra{\alpha;k\sb{1}a\sb{1},\cdots,k\sb{m}a\sb{m}} 
\smatrix
\ket{\beta;l\sb{1}b\sb{1},\cdots,l\sb{n}b\sb{n}}|\sb{\phi=0}
\nonumber\\
&=&
[ R\sp{a\sb{1}}(k\sb{1})  \cdots R\sp{a\sb{m}} (k\sb{m}) 
  R\sp{b\sb{1}}(-l\sb{1}) \cdots R\sp{b\sb{n}} (-l\sb{n}) ]\sb{{\rm S}} 
\bra{\alpha} \smatrix \ket{\beta}|\sb{\phi=0},
\label{eq214}
\end{eqnarray}
where $k\sb{i}~(l\sb{i})$ and $a\sb{i}~(b\sb{i})$ are 
the four-momentum and isospin indices of the outgoing (incoming) pions,
respectively, and $\alpha$ and $\beta$ represent the states of other particles.
Here, $[\cdots]\sb{{\rm S}}$ indicates 
that we take normalized symmetric permutations
of the functional derivative operators contained therein;
explicitly, we have
\begin{equation}
[{\cal D}\sb{1}\cdots{\cal D}\sb{n}]\sb{{\rm S}} =
\frac{1}{n!} \sum\sb{{\rm perms}} {\cal D}\sb{1}\cdots{\cal D}\sb{n}.
\label{eq215}
\end{equation}
This operation clearly shows the crossing symmetry in Eq.~(\ref{eq214}).
By using Eq.~(\ref{eq214}), together with Eq.~(\ref{eq24}),
scattering amplitudes can be expressed in terms of 
the Green functions of the operators 
$\hat{\cal O}=(\vectorjc{}{},\axialjc{}{},\fpi\hat{\sigma},\hat{\pi})$.
The {\chrf} takes the form of a functional derivative, 
and all constraints that stem from the broken chiral symmetry 
are contained in $R\sp{a}(k)$.
The {\chrf} can be understood as the following replacement of commutators:
\begin{equation}
[ a\sb{{\rm in}}\sp{a}(k), \ ] \rightarrow R\sp{a}(k),
\ \ \ \ 
[ \ , a\sb{{\rm in}}\sp{a\dag}(k) ] \rightarrow R\sp{a}(-k).
\label{eq216}
\end{equation}
For instance, the double commutator is replaced as
\begin{equation}
[ a\sb{{\rm in}}\sp{a}(k\sb{1}), 
[ \smatrix , a\sb{{\rm in}}\sp{b\dag}(k\sb{2}) ] ]
\rightarrow 
R\sp{b}(-k\sb{2}) R\sp{a}(k\sb{1}) \smatrix.
\label{eq217}
\end{equation}
\section{Interpolating pion field}
\label{sec3}
In contrast to $V\sp{a}\sb{\mu}$ and $A\sp{a}\sb{\mu}$,
which are identified with physical electroweak currents,
the operators $\hat{\cal O}$ are convention-dependent.
The interpolating pion field $\hat{\pi}\sp{a}$ 
is not uniquely determined: 
It is only required that the asymptotic condition
$\hat{\pi}\sp{a}(x) \rightarrow \pi\sp{a}\sb{{\rm in,out}}(x)\ 
(x\sp{0} \rightarrow \mp\infty)$
be satisfied.
Equations~(\ref{eq22}) and~(\ref{eq23}) represent 
an arbitrary way of dividing $V\sp{a}\sb{\mu}$ and $A\sp{a}\sb{\mu}$, 
respectively.
For the redefinition of the pion field
$\hat{\pi}\sp{a} \rightarrow \hat{\pi}\sp{\prime a}$,
with
\begin{equation}
\hat{\pi}\sp{\prime a}(x) =
\hat{\pi}\sp{a}(x) + \Lambda\sp{a}(x),
\label{eq31}
\end{equation}
$\vectorjc{a}{\mu}$ and $\axialjc{a}{\mu}$ also change in such a manner 
that Eqs.~(\ref{eq22}) and~(\ref{eq23}) remain invariant~\cite{Yam96}:
\begin{equation}
\vectorjc{\prime a}{\mu}(x) =
\vectorjc{a}{\mu}(x) - \fpi\ubexav{ac}{\mu} \Lambda\sp{c}(x),
\label{eq32}
\end{equation}
\begin{equation}
\axialjc{\prime a}{\mu}(x) =
\axialjc{a}{\mu}(x) + \fpi\nabla\sp{ac}\sb{\mu} \Lambda\sp{c}(x).
\label{eq33}
\end{equation}

The arbitrariness of the operators $\hat{\cal O}$ is translated into 
that of the extended S-matrix.
The redefined operators 
$\hat{\cal O}\sp{\prime} = 
(\vectorjc{\prime}{},\axialjc{\prime}{}, 
\fpi\hat{\sigma}\sp{\prime},\hat{\pi}\sp{\prime})$
are obtained in terms of an appropriately extended S-matrix 
$\smatrix\sp{\prime}[\phi]$, which
equals $\smatrix[\phi]$ at $\phi=0$ 
(i.e. $\smatrix|\sb{\phi=0}=\smatrix\sp{\prime}|\sb{\phi=0}\equiv\smatrixon$),
through functional derivatives as
\begin{equation}
\hat{\cal O}\sp{\prime}(x) =
-i\smatrix\sp{\prime\dag} \diff{\smatrix\sp{\prime}}{\phi(x)}.
\label{eq34}
\end{equation}
Therefore, a different way of extending the S-matrix $\smatrixon$ 
by introducing the external fields provides a different choice of 
the current and density operators.

The framework presented in Ref.~\citen{Yam96} only takes
account of the interpolating pion fields
given by the extended S-matrices satisfying the master equations.
Because the master equations fix the extended S-matrix up to 
a phase factor,\footnote{
A new extended S-matrix $\smatrix\sp{\prime}$ can be written as
$\smatrix\sp{\prime} = \smatrix \exp(if[\phi])$\cite{Yam96}.
Causality implies that the functional $f[\phi]$ has the form
\begin{equation*}
f[\phi] = \int d\sp{4}x P(x),
\end{equation*}
where $P(x)$ is a polynomial in $\phi(x)$ and its derivatives. 
[A constant term can be eliminated from $P(x)$ by imposing 
the normalization $\bra{0}\smatrixon\ket{0}=1$.]
Then, for \textit{any} $f[\phi]$ satisfying $T\sb{V,A}f[\phi]=0$,
the master equations (\ref{eq25}) and (\ref{eq26}) are invariant.
}
the arbitrariness in the choice of $\hat{\cal O}$ still exists.
However, as long as the extended S-matrices are introduced 
in such a manner that the master equations are satisfied, 
we can apply the {\chrf} to all of them.
The scattering amplitudes obtained from such extended S-matrices 
then satisfy the Ward identity given by the {\chrf},
and thus exhibit the same off-shell momentum dependence up to terms 
that are not fixed by the broken chiral symmetry.
(Of course, these amplitudes are identical on shell.)
A class of interpolating pion fields 
that is chosen to be consistent with the master equations
ensures the ``form invariance'' of the Ward identities.

We explain the situation described above more concretely 
by considering the $\pi N$ scattering as an example.
Using the {\chrf}~(\ref{eq214}), 
together with Eqs.~(\ref{eq21}) and~(\ref{eq24}),
we obtain
the $\pi\sp{a}(k)+N(p)\rightarrow\pi\sp{b}(k\sp{\prime})+N(p\sp{\prime})$ 
scattering amplitude $i{\cal T}\sb{\pi N}$ with two off-shell pions
(i.e.~$k\sp{2},k\sp{\prime 2}{\not =}\mpi\sp{2}$)
as
\begin{eqnarray}
i{\cal T}\sb{\pi N} 
&=&
[ R\sp{a}(-k) R\sp{b} (k\sp{\prime}) ]\sb{{\rm S}}
\bra{N(p\sp{\prime})} \smatrix \ket{N(p)}|\sb{\phi = 0}
\nonumber\\
&=&
i{\cal T}\sb{S} + i{\cal T}\sb{V} + i{\cal T}\sb{AA},
\label{eq35}
\end{eqnarray}
where
\begin{equation}
i{\cal T}\sb{S} =
-i\frac{k\sp{2}+k\sp{\prime 2}}{2\fpi} 
\delta\sp{ab} \bra{N(p\sp{\prime})} \hat{\sigma}(0) \ket{N(p)},
\label{eq36}
\end{equation}
\begin{equation}
i{\cal T}\sb{V} 
=
\frac{1}{2\fpi\sp{2}} (k+k\sp{\prime})\sp{\mu} \varepsilon\sp{abc}
\bra{N(p\sp{\prime})} \vectorjc{c}{\mu}(0) \ket{N(p)},
\label{eq37}
\end{equation}
and
\begin{equation}
i{\cal T}\sb{AA} =
-\frac{1}{\fpi\sp{2}}k\sp{\mu}k\sp{\prime \nu}
\int d\sp{4}x e\sp{ikx}
\bra{N(p\sp{\prime})} T\sp{\ast}(\axialjc{a}{\mu}(x)\axialjc{b}{\nu}(0)) \ket{N(p)},
\label{eq38}
\end{equation}
modulo $(2\pi)\sp{4}\delta\sp{(4)}(k+p-k\sp{\prime}-p\sp{\prime})$.
Now we redefine the operators as
$\hat{\cal O} \rightarrow \hat{\cal O}\sp{\prime}$,
which follows from the change of the extended S-matrix
$\hat{\cal S} \rightarrow \hat{\cal S}\sp{\prime}$.
Then, if $\hat{\cal S}\sp{\prime}$ satisfies the master equations, 
the {\chrf} can be applied to it.
The amplitude $i{\cal T}\sp{\prime}\sb{\pi N}$ 
obtained from $\hat{\cal S}\sp{\prime}$ 
takes the same form as $i{\cal T}\sb{\pi N}$, 
except that the Green functions of $\hat{\cal O}$
are replaced with those of $\hat{\cal O}\sp{\prime}$,
and thus
the overall momentum dependences of $i{\cal T}\sb{\pi N}$ and
$i{\cal T}\sp{\prime}\sb{\pi N}$ do not change.
The change of the amplitude resulting from 
the redefinition of the pion field only appears 
in the Green functions of $\hat{\cal O}$, 
which are not uniquely determined by the broken chiral symmetry.
In addition, the difference between the Green functions of 
$\hat{\cal O}$ and $\hat{\cal O}\sp{\prime}$ 
is only in some polynomials in the momenta,\footnote{
This results from the fact that
the phase $f[\phi]$ appearing in 
$\smatrix\sp{\prime}=\smatrix\exp(if[\phi])$
consists only of some polynomials in $\phi$ and their derivatives.
}
and this does not alter on-shell quantities 
that are determined by pole terms\cite{Yam96}.

The interpolating pion fields defined consistently with the 
master equations can be formally expressed as
\begin{equation}
\hat{\pi}\sp{a}(x) =
\hat{\pi}\sp{a}\sb{\text{PCAC}}(x) + F\sp{a}(\phi,\hat{\cal O}),
\label{eq39}
\end{equation}
where $\hat{\pi}\sp{a}\sb{\text{PCAC}}(x)$ is 
the so-called PCAC choice of the pion field
$\hat{\pi}\sb{\text{PCAC}}\sp{a} \equiv 
\partial\sp{\mu}A\sb{\mu}\sp{a}/(\fpi\mpi\sp{2})$,
and
$F\sp{a}(\phi,\hat{\cal O})$ is some function of $\phi$ and $\hat{\cal O}$
satisfying $F\sp{a}(\phi,\hat{\cal O})|\sb{\phi=0}=0$.
The asymptotic one-pion component is entirely included in 
$\hat{\pi}\sb{\text{PCAC}}\sp{a}$,
and therefore the second term on the r.h.s., $F\sp{a}(\phi,\hat{\cal O})$,
can generate only an off-shell effect.
Furthermore, $F\sp{a}(\phi,\hat{\cal O})$ is the only quantity 
that depends on the choice of the extended S-matrix,
$F\sp{a}(\phi,\hat{\cal O}) \rightarrow 
F\sp{a}(\phi,\hat{\cal O}\sp{\prime})$,
while $\hat{\pi}\sp{a}\sb{\text{PCAC}}$ 
is invariant under the redefinition of the pion field.
These facts imply that $F\sp{a}(\phi,\hat{\cal O})$ 
influences only the Green functions of $\hat{\cal O}$
in the Ward identity.
The overall kinematical dependence and symmetry structure 
of the off-shell Ward identity is determined by the first term in
Eq.~(\ref{eq39}), i.e. the PCAC pion field $\hat{\pi}\sp{a}\sb{\text{PCAC}}$.
Therefore, the class of interpolating pion fields
defined consistently with the master equations
and employed in the framework of Ref.~\citen{Yam96}
gives off-shell amplitudes satisfying Ward identities
of the same form as those obtained using the PCAC pion field.
\section{Extending the {\chrf} off mass shell}
\label{sec4}
The chiral Ward identity for the $\pi N$ scattering 
has also been derived by Weinberg,
using a different approach based on 
the chiral algebra satisfied by currents and 
density operators~\cite{Wei-70,Bro71}.
In that derivation, the PCAC choice of the pion field,
$\hat\pi\sp{a}\sb{\text{PCAC}}$, was employed to relate the $\pi N$ 
scattering amplitude
to the Ward identity among the Green functions of the currents and 
density operators.
If the prescription (\ref{eq214}) can be applied to 
the off-shell pions, the {\chrf} must reproduce 
Weinberg's formula, even 
off the mass shell, because the two approaches use the same definition
of the interpolating pion field.
However, 
for $i{\cal T}\sb{S}$ including the pion-nucleon $\sigma$-term,
Weinberg's formula gives 
\begin{equation}
i{\cal T}\sb{S} =
-i\frac{k\sp{2}+k\sp{\prime 2} - \mpi\sp{2}}{\fpi} 
\delta\sp{ab} \bra{N(p\sp{\prime})} \hat{\sigma}(0) \ket{N(p)},
\label{eq41}
\end{equation}
instead of Eq.~(\ref{eq36}).
This result indicates that a naive use of the {\chrf} (\ref{eq214})
proposed in Ref.~\citen{Yam96} 
does not correctly relate the \textit{off-shell} amplitudes 
to the Ward identities among
the Green functions of $\hat{\cal O}$; 
more specifically,
it does not do this in a manner
consistent with the master equations.

To make the {\chrf} applicable to the off-shell pions, 
first we note that the off-shell extrapolation
of scattering amplitudes is generally realized
through the LSZ reduction formula.
We therefore examine the derivation of {\chrf} starting from
the LSZ formalism.

We consider the commutation relations~(\ref{eq212}) 
on the basis of the LSZ formalism. 
From the asymptotic behavior of the interpolating pion field, we obtain
\begin{equation}
[ a\sb{{\rm in}}\sp{a}(k), \smatrix ] =
\int d\sp{4}x e\sp{+ikx} ( \Box + \mpi\sp{2} ) 
\diff{}{J\sp{a}(x)} \smatrix,
\label{eq42}
\end{equation}
\begin{equation}
[ \smatrix, a\sb{{\rm in}}\sp{a\dag}(k) ] =
\int d\sp{4}x e\sp{-ikx} ( \Box + \mpi\sp{2} ) 
\diff{}{J\sp{a}(x)} \smatrix,
\label{eq43}
\end{equation}
where we have used 
$a\sb{{\rm out}} = \smatrix\sp{\dag} a\sb{{\rm in}} \smatrix$
and Eq.~(\ref{eq21}).
Note that these exact relations are obtained independently of 
the master equations.
Once the commutation relations are rewritten in the form of 
Eqs.~(\ref{eq42}) and (\ref{eq43}),
the momentum $k$ can be analytically continued off the mass shell. 
Using these relations iteratively, 
we obtain the conventional LSZ reduction formula,
\begin{eqnarray}
&&
\bra{ \alpha; k\sb{1}a\sb{1}, \cdots, k\sb{m}a\sb{m}} 
\smatrix
\ket{ \beta;  l\sb{1}b\sb{1}, \cdots, l\sb{n}b\sb{n}}|\sb{\phi=0}
\nonumber\\
&=&
\left( 
\prod\sb{i=1}\sp{m} G(k\sb{i};x\sb{i}) 
\diff{}{J\sp{a\sb{i}}(x\sb{i})} 
\right)
\left( 
\prod\sb{j=1}\sp{n}
G(-l\sb{j};y\sb{j})
\diff{}{J\sp{b\sb{j}}(y\sb{j})} 
\right)
\bra{\alpha} \smatrix \ket{\beta}|\sb{\phi=0}
\nonumber\\
&=&
\left( \prod\sb{i=1}\sp{m} iG(k\sb{i};x\sb{i}) \right)
\left( \prod\sb{j=1}\sp{n} iG(-l\sb{j};y\sb{j}) \right)
\nonumber\\
&&
\times
\bra{\alpha}
\smatrixon
T\sp{\ast}
(
\hat{\pi}\sp{a\sb{1}}(x\sb{1}) \cdot\cdot \hat{\pi}\sp{a\sb{m}}(x\sb{m})
\hat{\pi}\sp{b\sb{1}}(y\sb{1}) \cdot\cdot \hat{\pi}\sp{b\sb{n}}(y\sb{n})
)
\ket{\beta},
\label{eq44}
\end{eqnarray}
where $G(k;x) = \int d\sp{4}x \exp(ikx)(\Box\sb{x}+\mpi\sp{2})$
and $\smatrixon = \smatrix|\sb{\phi=0}$, and 
we have used Eq.~(\ref{eq24}) in the last step.
This formula defines the off-shell extrapolation of the scattering amplitude.

We stress that owing to Eqs.~(\ref{eq28}) and~(\ref{eq213}), 
the \textit{identity} 
\begin{equation}
\int d\sp{4}x e\sp{+ikx} 
( \Box + \mpi\sp{2} ) \diff{}{J\sp{a}(x)}
= 
R\sp{a}(k) - T\sb{A}\sp{a}(k)
\label{eq45}
\end{equation}
exists among functional derivative operators,
where $T\sb{A}\sp{a}(k)$ is the Fourier transformation of $T\sb{A}\sp{a}(x)$.
Thus, we immediately find the following replacements for the commutators:
\begin{equation}
[ a\sb{in}\sp{a}(k), \ ] \rightarrow R\sp{a}(k) - T\sb{A}\sp{a}(k),
\label{eq46}
\end{equation}
\begin{equation}
[\ , a\sb{in}\sp{a\dag}(k)] \rightarrow R\sp{a}(-k) - T\sb{A}\sp{a}(-k).
\label{eq47}
\end{equation}
With these replacements, the scattering amplitude including
the $n$ incoming and $m$ outgoing (on-shell or off-shell) pions becomes
\begin{eqnarray}
&&
\bra{\alpha; k\sb{1}a\sb{1}, \cdots, k\sb{m}a\sb{m}} 
\smatrix
\ket{\beta;  l\sb{1}b\sb{1}, \cdots, l\sb{n}b\sb{n}}|\sb{\phi=0}
\nonumber\\
&=&
[ (R-T\sb{A})\sp{a\sb{1}}(k\sb{1})  \cdots (R-T\sb{A})\sp{a\sb{m}}(k\sb{m}) 
\nonumber\\
&&
\times
  (R-T\sb{A})\sp{b\sb{1}}(-l\sb{1}) \cdots (R-T\sb{A})\sp{b\sb{n}}(-l\sb{n}) 
]\sb{{\rm S}}
\bra{\alpha} \smatrix \ket{\beta}|\sb{\phi=0}
\nonumber\\
&=&
[ R\sp{a\sb{1}}(k\sb{1})  \cdots R\sp{a\sb{m}}(k\sb{m}) 
  R\sp{b\sb{1}}(-l\sb{1}) \cdots R\sp{b\sb{n}}(-l\sb{n}) ]\sb{{\rm S}} 
\bra{\alpha} \smatrix \ket{\beta}|\sb{\phi=0}
\nonumber\\
&&
+({\rm terms~including}\ T\sb{A}).
\label{eq48}
\end{eqnarray}
This is just another expression of the LSZ reduction formula.
Note here that we have not yet imposed any
constraints from the broken chiral symmetry.

The {\chrf} in our case is obtained as a combination of Eq.~(\ref{eq48}) 
and the master equation~(\ref{eq26}), which includes any effects
from the broken chiral symmetry.
To see our {\chrf} concretely, 
we consider the case $n+m=2$, i.e. 
that in which the sum of the incoming and outgoing pions is 2.
In this case, Eq.~(\ref{eq48}) gives
\begin{eqnarray}
&&
[ (R - T\sb{A})\sp{a} (R - T\sb{A})\sp{b} ]\sb{{\rm S}} 
\bra{\alpha} \smatrix \ket{\beta}|\sb{\phi=0}
\nonumber\\
&=&
  [ R\sp{a} R\sp{b} ]\sb{{\rm S}} 
\bra{\alpha} \smatrix \ket{\beta}|\sb{\phi=0}
+\left(
  [ T\sb{A}\sp{a} T\sb{A}\sp{b} ]\sb{{\rm S}}
- [ T\sb{A}\sp{a} R\sp{b} ]\sb{{\rm S}}
- [ R\sp{a} T\sb{A}\sp{b} ]\sb{{\rm S}}
\right)
\bra{\alpha} \smatrix \ket{\beta}|\sb{\phi=0},
\label{eq49}
\end{eqnarray}
where we have suppressed the momentum indices.
Note again that this is simply the LSZ reduction formula
expressed in terms of $R$ and $T\sb{A}$.
Next, we move all $T\sb{A}$ to the right of all $R$,
using the relation $T\sb{A}R = [ T\sb{A} , R ] + RT\sb{A}$,
and then we use the master equation~(\ref{eq26}).
As a result of this operation,
Eq.~(\ref{eq49}) becomes
\begin{eqnarray}
&&
[ (R - T\sb{A})\sp{a} (R - T\sb{A})\sp{b} ]\sb{{\rm S}} 
\bra{\alpha} \smatrix \ket{\beta}|\sb{\phi=0}
\nonumber\\
&=&
  [ R\sp{a} R\sp{b} ]\sb{{\rm S}} 
\bra{\alpha} \smatrix \ket{\beta}|\sb{\phi=0}
-\frac{1}{2}
\left(
  [ T\sb{A}\sp{a}, R\sp{b} ]
+ [ T\sb{A}\sp{b}, R\sp{a} ]
\right)
\bra{\alpha} \smatrix \ket{\beta}|\sb{\phi=0}.
\label{eq410}
\end{eqnarray}
This is the desired {\chrf} for the case $n+m=2$.
The above procedure is applied similarly for all 
values of $n$ and $m$.
The {\chrf} obtained in this manner is manifestly consistent with 
the LSZ reduction formula,
and there is no ambiguity in its derivation.

We note the presence of ``(terms including $T\sb{A}$)'' 
appearing in Eq.~(\ref{eq48}). 
In the following, we show that 
(i) such additional terms vanish if all pions are on shell, 
and therefore in that case our {\chrf} properly 
reduces to that proposed in Ref.~\citen{Yam96}, 
and that 
(ii) those terms are essential for deriving the correct
identities for the \textit{off-shell} amplitudes within the framework of
Ref.~\citen{Yam96}.

The assertion (i) can be readily demonstrated by considering 
the explicit expression of the commutation relation
between $T\sb{A}$ and $R$,
\begin{eqnarray}
[ T\sb{A}\sp{a}(k), R\sp{b}(k\sp{\prime}) ]
&=&
- \varepsilon\sp{abc}\frac{1}{\fpi\sp{2}}
  \int d\sp{4}x e\sp{i(k+k\sp{\prime})x}  
 T\sb{V}\sp{c}(x)
\nonumber\\
&&
- i\delta\sp{ab}(-k\sp{\prime 2}+\mpi\sp{2})
  (2\pi)\sp{4}\delta\sp{(4)}(k+k\sp{\prime})
\nonumber\\
&&
- \delta\sp{ab}(-k\sp{\prime 2}+\mpi\sp{2}) \frac{1}{\fpi\sp{2}} 
  \int d\sp{4}x e\sp{i(k+k\sp{\prime})x} \diff{}{s(x)}.
\label{eq411}
\end{eqnarray}
As found from Eqs.~(\ref{eq26}) and (\ref{eq411}), 
``(terms including $T\sb{A}$)'' in Eq.~(\ref{eq48}) 
is always proportional to $-k\sp{2}+\mpi\sp{2}$,
where $k$ is the four momentum of some external pion.
[By virtue of the symmetric permutation~(\ref{eq215}), 
the contribution from 
the first term in Eq.~(\ref{eq411}) involving the antisymmetric tensor 
$\varepsilon\sp{abc}$ vanishes.]
As a result, our {\chrf}, i.e. Eq.~(\ref{eq48}) 
together with Eq.~(\ref{eq26}), reduces to 
the {\chrf} proposed in Ref.~\citen{Yam96} if all pions are on shell.

To verify the assertion (ii), we return to the investigation of 
the $\pi N$ scattering.
With the one-nucleon states for $\bra{\alpha}$ and $\ket{\beta}$,
using Eqs.~(\ref{eq21}) and (\ref{eq411}), 
it can be shown that Eq.~(\ref{eq410}) becomes
\begin{eqnarray}
&&
[ (R - T\sb{A})\sp{a}(-k) (R - T\sb{A})\sp{b}(k\sp{\prime}) ]\sb{{\rm S}} 
\bra{N(p\sp{\prime})} \smatrix \ket{N(p)}|\sb{\phi=0}
\nonumber\\
&=&
[ R\sp{a}(-k) R\sp{b}(k\sp{\prime}) ]\sb{{\rm S}} 
\bra{N(p\sp{\prime})} \smatrix \ket{N(p)}|\sb{\phi=0}
\nonumber\\
&&
-i\frac{\delta\sp{ab}}{\fpi}
\left( \frac{k\sp{2} + k\sp{\prime 2}}{2} -\mpi\sp{2} \right)
\bra{N(p\sp{\prime})} \hat{\sigma}(0) \ket{N(p)},
\label{eq412}
\end{eqnarray}
modulo $(2\pi)\sp{4}\delta\sp{(4)}(k+p-k\sp{\prime}-p\sp{\prime})$,
where we drop the non-scattering parts arising from the second term
on the r.h.s. of Eq.~(\ref{eq411}).
Now we obtain the additional contribution to Eq.~(\ref{eq35}) 
arising from ``(terms including $T\sb{A}$)''.
Our result is identical with Weinberg's formula for the $\pi N$ scattering,
even off the mass shell, as should be the case.
We can thus conclude that our {\chrf}, being manifestly
consistent with the LSZ reduction formula, is
a proper off-shell extension of the on-shell {\chrf} 
presented in Ref.~\citen{Yam96}. 

Before closing this section, several comments are in order:
\begin{itemize}
\item
A naive application of the replacement of
the single commutators expressed in Eq.~(\ref{eq216}) 
to multi-commutator cases results in
the nontrivial replacement of the functional derivative operator
$\int d\sp{4}x \exp(ikx) ( \Box + \mpi\sp{2} )(\delta / \delta J\sp{a}) 
\rightarrow R\sp{a}(k)$.
Such replacement is consistent with the LSZ reduction formula
for on-shell amplitudes, but it is incompatible off the mass shell,
because $R$ and $T\sb{A}$ do not commute,
and thus ``(terms including $T\sb{A}$)'' does not vanish.
This is the reason that the {\chrf} proposed in Ref.~\citen{Yam96}
cannot be naively applied to off-shell pions.
\item
Practically, the {\chrf} formulated in Ref.~\citen{Yam96} gives 
the same off-shell result
as ours for single pion emission and absorption processes
[see Eqs.~(6.4) and (6.5) in Ref.~\citen{Yam96}]. 
According to our results, however, it seems to be a rather special case,
because in this case, ``(terms including $T\sb{A}$)''
survives in neither the on-shell nor off-shell case:
$\bra{\alpha}[a\sb{\text{in}}\sp{a}(k),\smatrix]\ket{\beta}
=(R-T\sb{A})\sp{a}(k)\bra{\alpha}\smatrix \ket{\beta}
=R\sp{a}(k)\bra{\alpha}\smatrix\ket{\beta}$.
\item
In principle, the {\chrf} presented in this paper is
applicable for any value of (on-shell or off-shell) pion momenta.
This is due to the fact that
the master equations are exact formulas essentially equivalent to 
the current conservation law or Noether's theorem.
The nature of specific points, such as the chiral limit and 
the soft pion limit, is relevant only 
if extra assumptions are introduced, for example, 
the possibility of a loop expansion around those points\cite{Yam96}.
\end{itemize}
\section{Summary}
\label{sec5}
We have investigated the off-shell structure of the {\chrf}
because of its importance in actual calculations and in the 
pursuit of a deeper
understanding of the theoretical framework developed 
in Ref.~\citen{Yam96}.
We have seen that the on-shell {\chrf} proposed in Ref.~\citen{Yam96}
cannot be naively applied to off-shell pions.
We then found its proper extension off the mass shell.
This was achieved by reconstructing the {\chrf}
in a form manifestly consistent with 
the conventional LSZ reduction formula.

The analysis employed in this work can be 
readily applied to the three flavor case studied in 
Refs.~\citen{Lee99} and \citen{Lee98}.
\section*{Acknowledgments}
The author would like to thank Dr. Masaki Arima 
for discussions and his helpful advice in completing the manuscript.
The author would also like to thank Dr. Toru Sato
for valuable comments.
This work was supported by Research Fellowships 
of the Japan Society for the Promotion of Science (JSPS)
for Young Scientists.
%
%
%
%
%
%
%


\begin{thebibliography}{14}
\expandafter\ifx\csname url\endcsname\relax
  \def\url#1{\texttt{#1}}\fi
\expandafter\ifx\csname urlprefix\endcsname\relax\def\urlprefix{URL }\fi
\bibitem{Yam96}
H.~Yamagishi and I.~Zahed, \ANN{247,1996,292}.

\bibitem{Vel66}
M.~Veltman, \PRL{17,1966,553}.

\bibitem{Bell67}
J.~S.~Bell, \JL{Nuovo~Cim. (Ser.~X) A,50,1967,129}.

\bibitem{Ste97}
J.~V. Steele, H.~Yamagishi and I.~Zahed, \NPA{615,1997,305}.

\bibitem{Lee99}
C.-H. Lee, H.~Yamagishi and I.~Zahed, \NPA{653,1999,185}.

\bibitem{Kam05-1}
H.~Kamano, M.~Morishita and M.~Arima, \PRC{71,2005,045201}.

\bibitem{Ste96}
J.~V. Steele, H.~Yamagishi and I.~Zahed, \PLB{384,1996,255}.

\bibitem{Ste97-2}
J.~V. Steele, H.~Yamagishi and I.~Zahed, \PRD{56,1997,5605}.

\bibitem{Lee98}
C.-H. Lee, H.~Yamagishi and I.~Zahed, \PRC{58,1998,2899}.

\bibitem{Dus06}
K.~Dusling, D.~Teaney and I.~Zahed, nucl-th/0604071.

\bibitem{Bog}
N.~N. Bogoliubov and D.~V. Shirkov, 
\textit{Introduction to the Theory of Quantized Fields}, 3rd ed.
(Wiley, New York, 1980).

\bibitem{Wei-70}
S.~Weinberg, 
\textit{Lectures on Elementary Particles and Quantum Field Theory}, 
S.~Deser, M.~Grisaru and H.~Pendleton, Vol.~1
(MIT Press, 1970).

\bibitem{Bro71}
L.~S. Brown, W.~J. Pardee and R.~D. Peccei, \PRD{4,1971,2801}.


\end{thebibliography}
\end{document}